\documentclass[12pt]{article}
\usepackage{times}
\usepackage{comment}
\usepackage{sidecap}

\usepackage{hyperref}
\hypersetup{
    colorlinks=true,
    linkcolor=blue,
    filecolor=magenta,      
    urlcolor=blue,
    citecolor=blue,
}

\usepackage{url}

\sidecaptionvpos{figure}{t}
\usepackage{geometry}
\usepackage{graphicx, caption}
\usepackage[utf8]{inputenc}
\usepackage[round]{natbib} 
\usepackage{enumitem,amssymb}
\usepackage{ragged2e}
\newlist{thematic}{itemize}{8}
\setlist[thematic]{label=$\square$}
\usepackage{pifont}

\def\Msun{M_\odot}
\usepackage{amsmath} 
\newcommand{\angstrom}{\textup{\AA}}

\begin{document}

\def\aj{{AJ}}                   
\def\actaa{\ref@jnl{Acta Astron.}}      
\def\araa{\ref@jnl{ARA\&A}}             
\def\apj{{ApJ}}                 
\def\apjl{{ApJ}}                
\def\apjs{{ApJS}}               
\def\ao{\ref@jnl{Appl.~Opt.}}           
\def\apss{\ref@jnl{Ap\&SS}}             
\def\aap{{A\&A}}                
\def\aapr{\ref@jnl{A\&A~Rev.}}          
\def\aaps{\ref@jnl{A\&AS}}              
\def\azh{\ref@jnl{AZh}}                 
\def\baas{\ref@jnl{BAAS}}               
\def\bac{\ref@jnl{Bull. astr. Inst. Czechosl.}}
\def\caa{\ref@jnl{Chinese Astron. Astrophys.}}
\def\cjaa{\ref@jnl{Chinese J. Astron. Astrophys.}}
\def\icarus{\ref@jnl{Icarus}}           
\def\jcap{\ref@jnl{J. Cosmology Astropart. Phys.}}
\def\jrasc{\ref@jnl{JRASC}}             
\def\memras{\ref@jnl{MmRAS}}            
\def\mnras{{MNRAS}}             
\def\na{\ref@jnl{New A}}                
\def\nar{\ref@jnl{New A Rev.}}          
\def\pra{\ref@jnl{Phys.~Rev.~A}}        
\def\prb{\ref@jnl{Phys.~Rev.~B}}        
\def\prc{\ref@jnl{Phys.~Rev.~C}}        
\def\prd{\ref@jnl{Phys.~Rev.~D}}        
\def\pre{\ref@jnl{Phys.~Rev.~E}}        
\def\prl{\ref@jnl{Phys.~Rev.~Lett.}}    
\def\pasa{\ref@jnl{PASA}}               
\def\pasp{\ref@jnl{PASP}}               
\def\pasj{\ref@jnl{PASJ}}               
\def\rmxaa{\ref@jnl{Rev. Mexicana Astron. Astrofis.}}%
\def\qjras{\ref@jnl{QJRAS}}             
\def\skytel{\ref@jnl{S\&T}}             
\def\solphys{\ref@jnl{Sol.~Phys.}}      
\def\sovast{\ref@jnl{Soviet~Ast.}}      
\def\ssr{\ref@jnl{Space~Sci.~Rev.}}     
\def\zap{\ref@jnl{ZAp}}                 
\def\nat{{Nature}}              
\def\iaucirc{\ref@jnl{IAU~Circ.}}       
\def\aplett{\ref@jnl{Astrophys.~Lett.}} 
\def\apspr{\ref@jnl{Astrophys.~Space~Phys.~Res.}}
\def\bain{\ref@jnl{Bull.~Astron.~Inst.~Netherlands}} 
\def\fcp{\ref@jnl{Fund.~Cosmic~Phys.}}  
\def\gca{\ref@jnl{Geochim.~Cosmochim.~Acta}}   
\def\grl{\ref@jnl{Geophys.~Res.~Lett.}} 
\def\jcp{\ref@jnl{J.~Chem.~Phys.}}      
\def\jgr{\ref@jnl{J.~Geophys.~Res.}}    
\def\jqsrt{\ref@jnl{J.~Quant.~Spec.~Radiat.~Transf.}}
\def\memsai{\ref@jnl{Mem.~Soc.~Astron.~Italiana}}
\def\nphysa{\ref@jnl{Nucl.~Phys.~A}}   
\def\physrep{{Phys.~Rep.}}   
\def\physscr{\ref@jnl{Phys.~Scr}}   
\def\planss{\ref@jnl{Planet.~Space~Sci.}}   
\def\procspie{\ref@jnl{Proc.~SPIE}}   

\let\astap=\aap
\let\apjlett=\apjl
\let\apjsupp=\apjs
\let\applopt=\ao

\begin{flushleft}
\Large
\begin{center}
A White Paper for the Astro2020 Decadal Survey
\end{center}

\LARGE
\begin{center}
\textbf{Detecting the Birth of Supermassive Black Holes} \\ \textbf{Formed from Heavy Seeds}
\end{center}
\normalsize

\noindent \textbf{Thematic Area:} Galaxy Evolution, Multi-Messenger Astronomy and Astrophysics, Formation and Evolution of Compact Objects, Cosmology and Fundamental Physics \linebreak
  
\textbf{Principal Author:}

Name: Fabio Pacucci
 \linebreak						
Institution: Kapteyn Astronomical Institute, Yale University
 \linebreak
Email: fabio.pacucci@yale.edu
 \linebreak
Phone: (203)298-2478
 \linebreak
 
\textbf{Co-authors:}\\ 
Vivienne Baldassare$^1$, Nico Cappelluti$^2$, Xiaohui Fan$^3$, Andrea Ferrara$^4$, Zoltan Haiman$^5$, Priyamvada Natarajan$^1$, Feryal Ozel$^3$, Raffaella Schneider$^6$, Grant R. Tremblay$^7$, Megan C. Urry$^1$, Rosa Valiante$^8$, Alexey Vikhlinin$^7$, Marta Volonteri$^9$
\linebreak

$^1$Yale University,
$^2$University of Miami,
$^3$University of Arizona,
$^4$Scuola Normale Superiore,
$^5$Columbia University, 
$^6$Sapienza Universit\`a di Roma,
$^7$Center for Astrophysics $|$ Harvard \& Smithsonian,
$^8$INAF - Roma,
$^9$Institut d'Astrophysique de Paris

\vspace{-0.4cm}
\end{flushleft}

\begin{figure}[h]
\hspace{-0.5cm}
\begin{center}
\includegraphics[angle=0,width=0.68\textwidth]{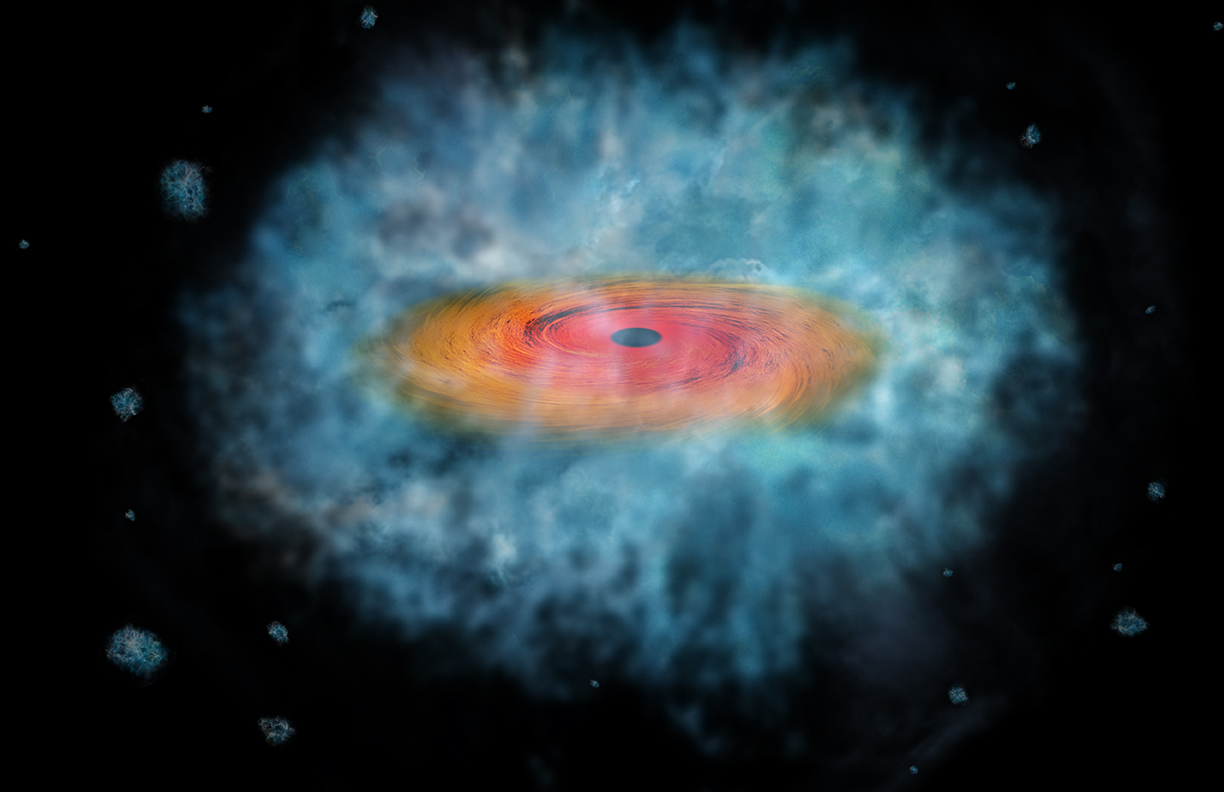}
\captionsetup{labelformat=empty}
\caption{\textit{Artistic representation of a heavy black hole seed, formed in the early Universe. Despite numerous theoretical and observational efforts to observe the birth of the first population of black holes, thus far we are still lacking a confirmed detection. The formation of these objects would be among the most spectacular events in the history of the Universe. (Credit: NASA/CXC/M. Weiss)}}
\label{fig:intro}
\end{center}
\end{figure}

\thispagestyle{empty}

\pagebreak
\clearpage
\setcounter{page}{1}
\setcounter{figure}{0}

\begin{center}
\textbf{Introduction}\\
\end{center}
\vspace{-0.3cm}
The dawn of the first black holes (and stars) occurred $\sim 100 \, \mathrm{Myr}$ after the Big Bang (\citealt{BL01}). It is very remarkable that numerous observations in the past two decades have shown the presence of Super-Massive Black Holes (SMBHs, with masses $10^{9-10} \, \mathrm{M_{\odot}}$) less than $700 \, \mathrm{Myr}$ later (e.g., \citealt{Fan_2006, Mortlock_2011, Wu_2015, Banados_2018}). A ``seed" is the original black hole that, growing via gas accretion and mergers, generates a SMBH. Seeds are categorized in light ($\lesssim 10^2 \, \mathrm{M_{\odot}}$, formed as stellar remnants) and heavy ($\sim 10^4-10^6 \, \mathrm{M_{\odot}}$). Heavy objects formed by the direct collapse of primordial gas clouds are named Direct Collapse Black Holes (DCBHs, \citealt{Haehnelt_1993, Bromm_Loeb_2003, Lodato_Natarajan_2006}).

It is challenging to grow a black hole from a light seed in time to match the observations of SMBHs in the early Universe (\citealt{Haiman_Loeb_2001, Haiman_2004}, see also reviews by \citealt{Volonteri_Bellovary_2012, Haiman_2013, Woods_2018}). Possible solutions to decrease the growth time are: (i) start the growth from heavy seeds \citep{Bromm_Loeb_2003}, and (ii) allow extremely large accretion rates in the high-$z$ Universe \citep{Begelman_1979, Volonteri_2005, Pacucci_2015, Inayoshi_2016, Pezzulli_2016}.
Notwithstanding several efforts in theoretical predictions and observations (e.g., \citealt{Sobral_2015, Pallottini_Pacucci_2015_CR7, PFVD_2015, Valiante_2018, Valiante_2018_b}), thus far there is no confirmed detection of any black hole seed. DCBHs may be relatively common in the early universe, with recent work suggesting a comoving number density of $\sim 10^{-6}\, \mathrm{Mpc^{-3}}$ and as high as $\sim 10^{-3}\, \mathrm{Mpc^{-3}}$ in dense regions \citep{Wise_2019}.
Previous studies which also explored the issue of DCBH formation include \cite{Yoshida_2003, Visbal_2014_DCBH, Chon_2016, Maio_2018, Inayoshi_2018}. These theoretical findings highlight the great potential relevance of this formation channel.
\vspace{0.2cm}

\begin{figure}[h]
  \centering
   \includegraphics[angle=0,width=0.90\textwidth]{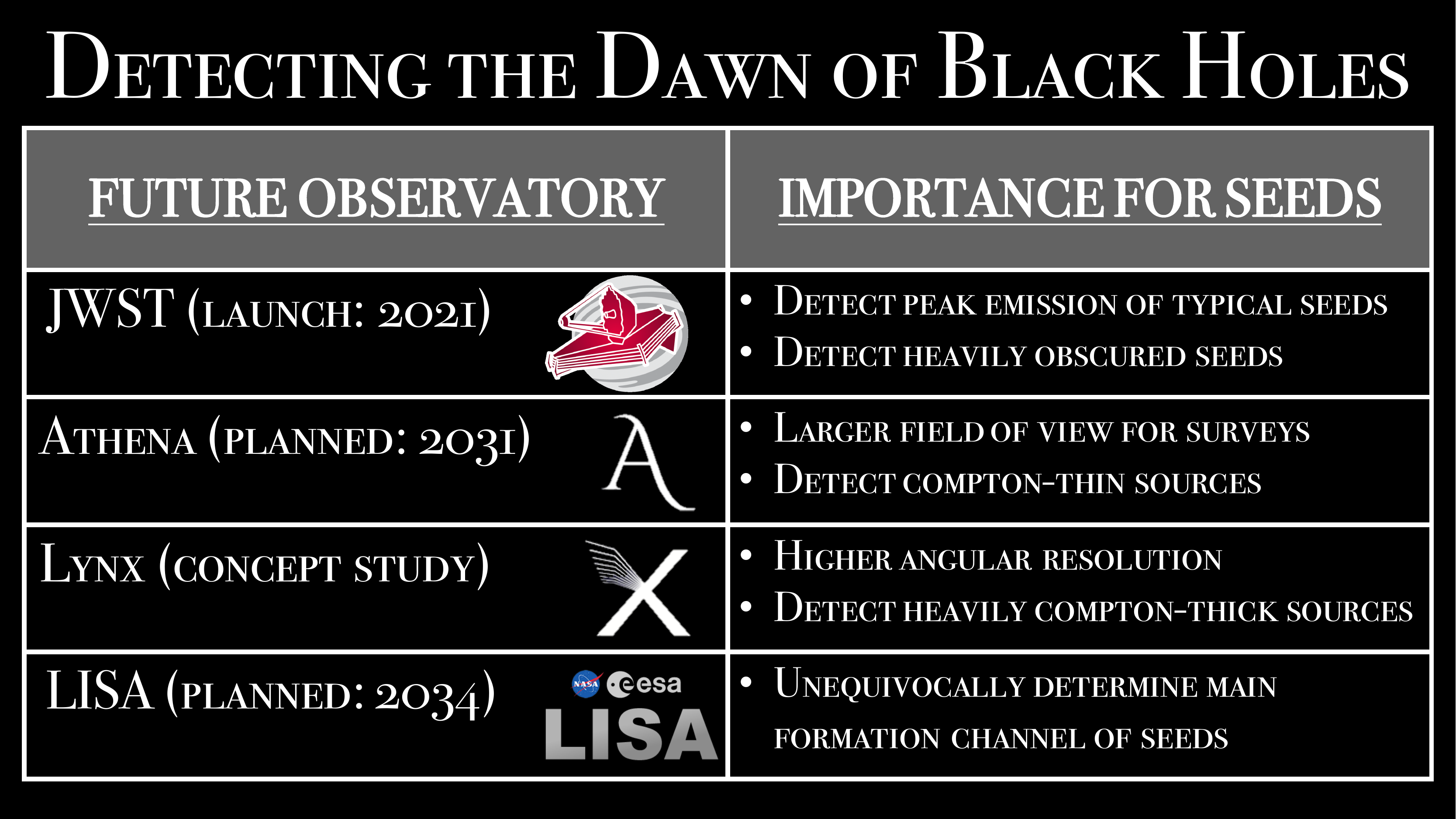}
\caption{Overview of future observatories that will detect the dawn of black holes.}
\vspace{-0.2cm}
\label{fig:schematic}
\end{figure}

\vspace{0.2cm}

\textbf{Shedding light on the dawn of black holes will be one of the key tasks that the astronomical community will focus on in the next decade.} The unknowns in this field are several and largely unconstrained.
\textit{What is the main formation channel? Assuming that both heavy and light seeds were formed, what is their typical formation ratio? What is the peak redshift of their formation?}
Investigating the dawn of black holes will have crucial consequences on the theory of galaxy formation/evolution and on gravitational wave astronomy. A better understanding of the initial conditions of this high-$z$ population will provide fundamental clues on its evolution at lower redshifts, down to the local Universe around us. 
In fact: (i) there is a tight connection between some properties of the host galaxy and the mass of the SMBH at its center (e.g., \citealt{Kormendy_Ho_2013}), and (ii) the progenitors of the merging black holes that we observe via gravitational waves could have formed as high-$z$ light seeds (e.g., \citealt{Kinugawa_2014}).

The formation of SMBHs by the DCBH scenario at $z \gtrsim 10$ is very appealing on many grounds (e.g., \citealt{Oh_Haiman_2002,Bromm_Loeb_2003, Lodato_Natarajan_2006, PFVD_2015}). Direct collapse of a gas cloud onto a $10^{4-6}$ black hole would be among the most spectacular events in the history of the Universe. \textbf{In this white paper we address the question of what capabilities are required to identify and study SMBHs formed by heavy seeds in the early Universe.} On similar topics, see the white papers by \cite{Natarajan_WP,Haiman_WP}.

In the electromagnetic spectrum, infrared and X-ray observations offer the best chances to investigate the dawn of black holes. In fact, while infrared wavelengths probe the spectral region of highest emission, X-ray photons are able to escape from the extremely large column densities that their hosts are predicted to have (e.g., \citealt{Pacucci_2015}). 
Future observatories in both spectral ranges, like the James Webb Space Telescope (\href{https://www.jwst.nasa.gov/}{\textit{JWST}}), \href{https://www.the-athena-x-ray-observatory.eu/}{\textit{Athena}} and the proposed \href{https://wwwastro.msfc.nasa.gov/lynx/}{\textit{Lynx}}, will certainly play a major role in unraveling the dawn of black holes (see Fig. \ref{fig:schematic}).
The \textit{JWST}, with its impressive angular resolution and a light-collecting area seven times larger than the Hubble Space Telescope (\textit{HST}), will observe the infrared sky farther than ever before. \textit{Athena}, with its large field of view, will be fundamental for X-ray surveys.
\textit{Lynx}, with excellent angular resolution, high throughput and spectral resolution for point-like and extended sources, will collect X-ray photons from the most obscured accreting sources in the high-$z$ Universe.

\begin{center}
\textbf{How can we identify $z \gtrsim 10$ heavy seeds?}\\
\end{center}
\vspace{-0.3cm}
Currently we probe only the most luminous high-$z$ black holes: $\sim 10^{9-10} \, \mathrm{\Msun}$ objects at $z \sim 6-7$ (\citealt{Fan_2006, Mortlock_2011, Banados_2018}). This is clearly the tip of the iceberg of their mass distribution (see e.g. the SHELLQs survey, \citealt{Subaru_2018}).
Upcoming facilities will revolutionize our view of the early Universe, by probing mass scales $\lesssim 10^6 \, \mathrm{\Msun}$ (\citealt{PFVD_2015, Woods_2018}). To exemplify the extent of the observational revolution that we are about to witness, NIRCam onboard the \textit{JWST} will reach $m=30.5$ at $5\sigma$ with an exposure of $\sim 88 \, \mathrm{hr}$ \citep{Finkelstein_2015}.  Depending on the models and on the brightness of the host galaxy, this will enable the detection of objects of $\sim 10^{5-6} \, \mathrm{\Msun}$ at $z \gtrsim 10$. These capabilities will open up, for the first time in history, the window to the dawn of black holes.

In order to observationally identify heavy seeds it is thus crucially important to understand the observational signatures that we are seeking. \textbf{To obtain an unequivocal detection of heavy seeds we need to probe mass scales of $\sim 10^{5-6} \, \mathrm{\Msun}$ at redshift $z \gtrsim 10$.} Observing them early in their evolution (i.e., at birth or soon after) is crucial: once a black hole has evolved from its original seed, the initial conditions are rapidly deleted and become undetectable \citep{Valiante_2018}.

The observational methodologies proposed can be divided in \textit{direct} and \textit{indirect}. A direct method affirms, within some error margin, whether a source is a heavy black hole seed or not. An indirect method, instead, looks at a group of objects and infer whether it is likely that at least a fraction of them originated from heavy seeds. In this white paper we focus on direct methods. See \cite{Haiman_WP} for a review of indirect methods.

\begin{figure}
\centering
    \begin{minipage}{0.50\textwidth}
  \centering
  \includegraphics[width=1.05\linewidth]{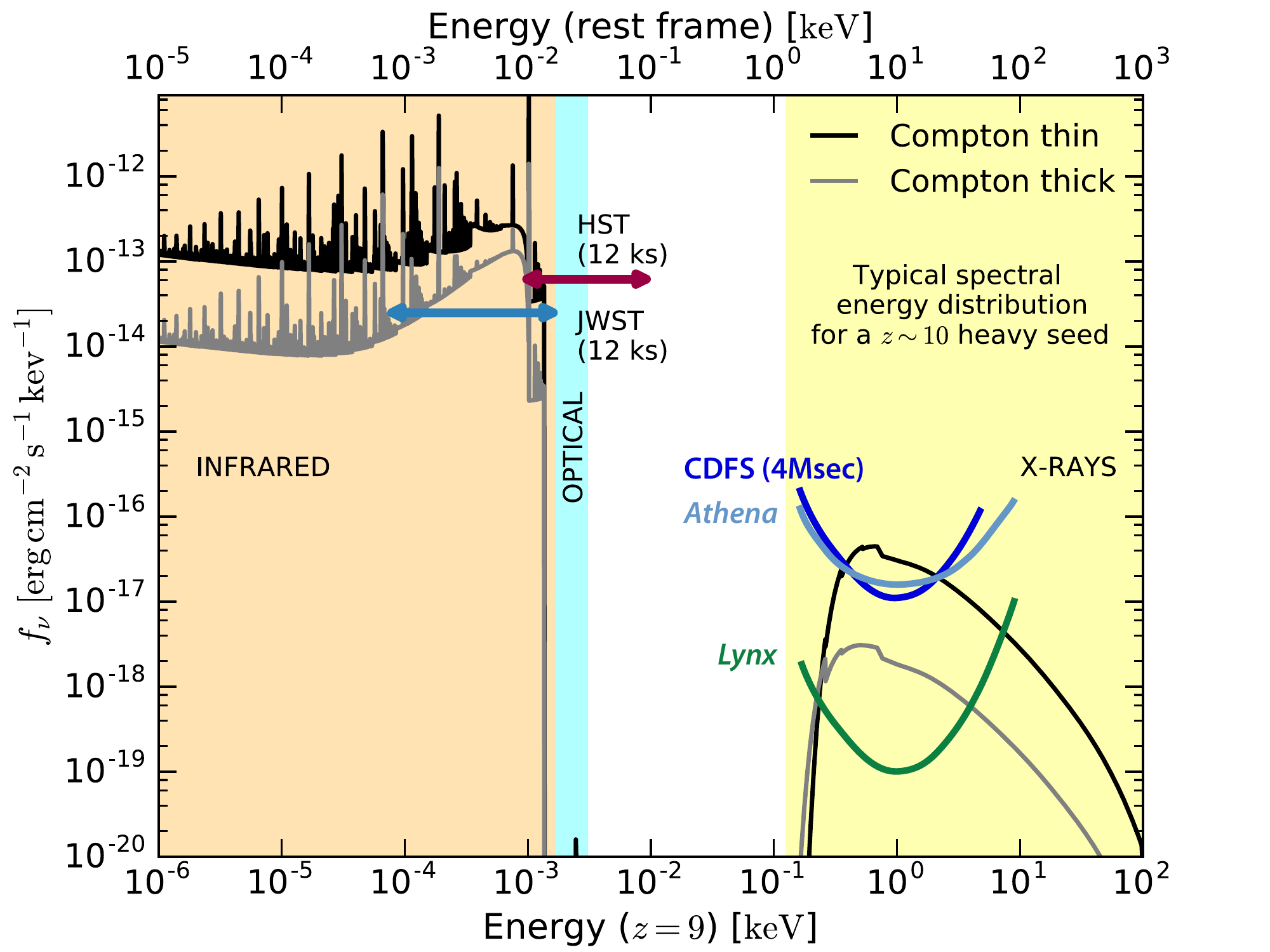}
 \end{minipage}\hfill
    \begin{minipage}{0.50\textwidth}
  \centering
  \includegraphics[width=1.05\linewidth]{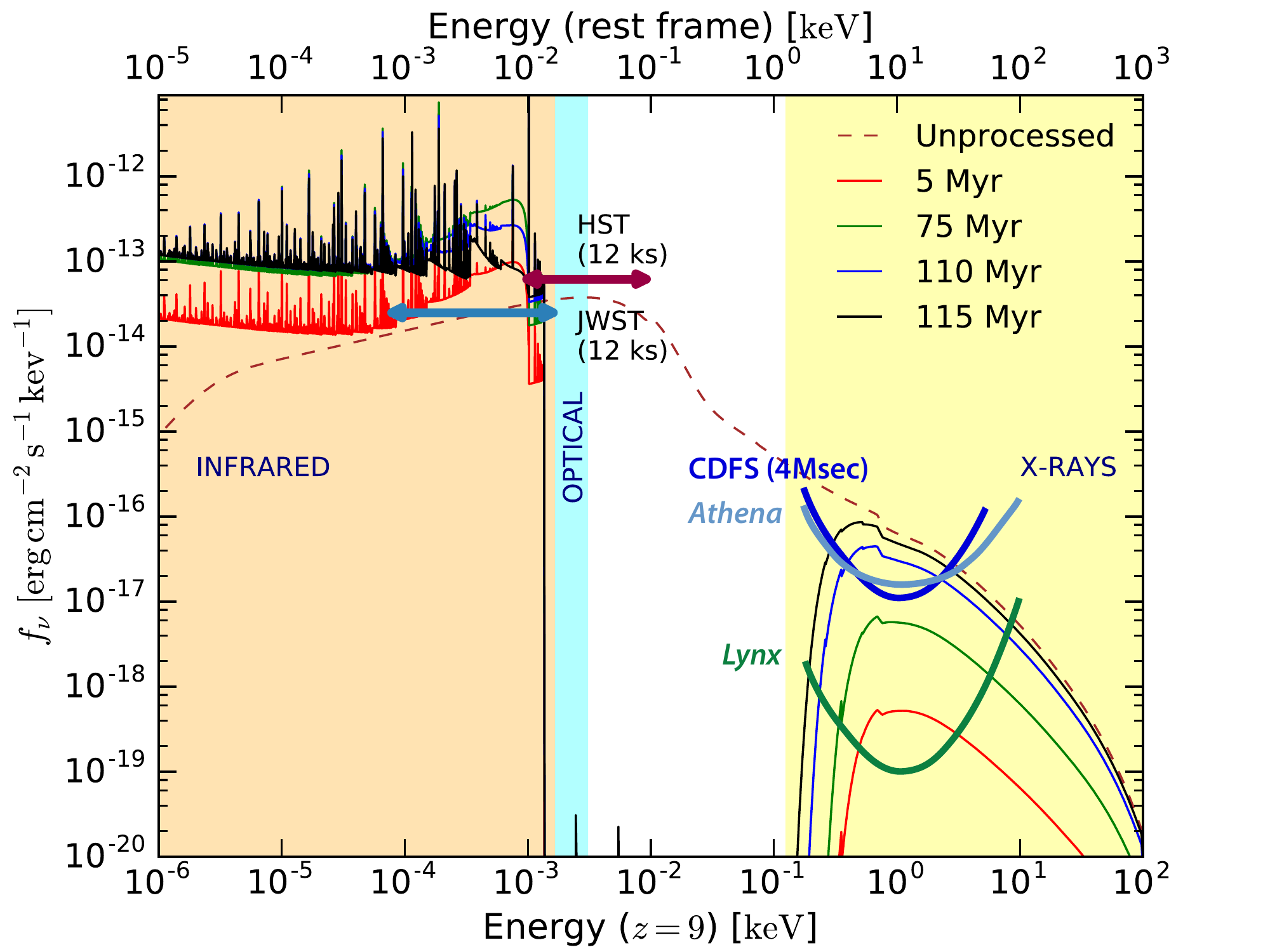}
\end{minipage}\hfill
\caption{Spectral predictions for a typical heavy seed. The flux limits for future/proposed (\textit{JWST}, \textit{Athena}, \textit{Lynx}) and current (\textit{HST}, \textit{Chandra}) observatories are shown. \textbf{Left:} Spectral predictions are shown at peak infrared emission and in the Compton thin and Compton thick cases.  \textbf{Right:} Spectral predictions are shown at different times during the seed evolution. The unprocessed spectrum refers to the radiation emitted by the central source and not processed by the host galaxy. Adapted from \citealt{PFVD_2015}.}
\label{fig:spectrum}
\end{figure}

The next generation of telescopes will provide an unprecedented number of high-quality spectra. Thus, it is important to understand the spectral signatures of heavy seeds.
\cite{PFVD_2015} presented the first, accurate study of spectral templates (continuum + lines) for heavy seeds, spanning from the sub-mm to the X-ray (see Fig. \ref{fig:spectrum}). 
For a typical heavy seed, buried under very large absorbing column densities, the emission occurs in the observed infrared-submm ($1-1000 \, \mathrm{\mu m}$) and X-ray ($0.1-100 \, \mathrm{keV}$) bands. These sources feature a very steep spectrum in the infrared, because much of the radiation emitted by the central source is reprocessed at lower energies by the intervening matter \citep{Pacucci_2016}. 
At the fiducial redshift $z = 10$, the signal generated by a heavy seeds will be easily detectable by the \textit{JWST} at a mass scale $\sim 10^5-10^6 \, \mathrm{\Msun}$, while \textit{Lynx} will go down to $\sim 10^4 \, \mathrm{\Msun}$; \textit{Athena} will be able to detect these sources down to a mass scale $\sim 10^6 \, \mathrm{\Msun}$. Thanks to its higher angular resolution, images from \textit{Lynx} will also be affected by a lower source confusion due to foreground objects. All the aforementioned estimates assume a Compton-thin irradiation scenario.
In fact, as shown in several studies (e.g., \citealt{Pacucci_2017_CR7, Valiante_2018_b}) the spectrum depends on multiple parameters: (i) black hole mass, (ii) column density and gas metallicity of the host, (iii) presence of stars. \cite{PFVD_2015, Natarajan_2017} indicate that heavy seeds will be primary targets for all these upcoming facilities.

Additional observational signatures of $z \gtrsim 10$ heavy seeds may come from the very large absorbing column density of their host galaxies (e.g., \citealt{Pacucci_2015, Latif_2013, Begelman_Volonteri_2017}). In fact, column densities comparable to or well in excess of the Compton-thick limit ($N_H \sim 1.5 \times 10^{24} \, \mathrm{cm^{-2}}$) may be reached during the formation of the seed, or shortly after for periods of $\sim 100 \, \mathrm{Myr}$ \citep{PVF_2015}. The effects of large absorbing column densities are: (i) X-ray fluxes in the soft and in the hard bands ($0.5-10 \, \mathrm{keV}$) are significantly reduced by factors up to $\sim 100$ for extreme values of the column density ($N_H \gtrsim 10^{25} \, \mathrm{cm^{-2}}$), and (ii) X-ray photons are re-emitted in the infrared bands, due to Auger-like cascade effects \citep{PFVD_2015}. In summary, the increase in column density to values $N_H \gtrsim 1.5 \times 10^{24} \, \mathrm{cm^{-2}}$ causes a decrease in the X-ray emission and an increase in the infrared emission. Supported by radiation-hydrodynamical simulations, \cite{PFVD_2015, Natarajan_2017} suggest that extremely absorbed heavy seeds will have a ratio of X-ray flux to optical flux $F_X/F_{444 \, \mathrm{\angstrom}} \gg 1$, and a ratio of X-ray flux to infrared flux $F_X/F_{1 \, \mathrm{\mu m}} \ll 1$.

Additional probes are available if future instruments will detect not only the central black hole, but also its host galaxy. For instance, \cite{Agarwal_2013, Visbal_Haiman_2018}, utilizing a cosmological N-body simulation, show that before the $\sim 10^5 \, \mathrm{\Msun}$ DCBH at $z \gtrsim 10$ grows to $\sim 10^6 \, \mathrm{\Msun}$, it will have a black hole mass to halo mass ratio larger than expected for remnants of Pop III stars, grown to the same mass. Thus, a combination of infrared and X-ray observations will be able to distinguish high-$z$ DCBHs from lighter seeds due to this peculiarly large ratio.

\vspace{0.1cm}

\begin{center}
\textbf{What observational capabilities do we need to detect $z \gtrsim 10$ heavy seeds?}\\
\end{center}
\vspace{-0.3cm}
Any effort to detect $z \gtrsim 10$ heavy seeds will likely need a pre-selection of sources with observational properties meeting some criteria. In this regard, a blind X-ray survey will play a fundamental role.
The final confirmation of the heavy seed nature of a $z \gtrsim 10$ source will eventually need a high-resolution spectrum showing high-ionization lines and no metal lines. 
\cite{Pacucci_2016} introduced a photometric selection criteria for heavy seeds, leading to the first identification of $z > 6$ candidates. The criteria arise from the observation that the infrared spectral energy distribution of seeds is predicted to be significantly steeper when compared to other high-$z$ sources. Other works extended this study \citep{Natarajan_2017, Volonteri_2017}. For instance, \cite{Valiante_2018_b} introduced a careful modeling of the metallicity and dust evolution of the hosts. Their improved selection method employs a combined analysis of near-infrared colours, infrared excess, and ultraviolet continuum slopes to distinguish host galaxies with growing heavy seeds from starburst-dominated systems in \textit{JWST} surveys.
The search for heavy seeds thus far is limited to the GOODS-S field, spanning $\sim 800 \, \mathrm{arcmin^2}$, leading to 2 candidates and the prediction of finding $\lesssim 10$ \citep{Natarajan_2017}. Applying the same criteria to larger surveys ($\gtrsim 1 \, \mathrm{deg^{-2}}$) will allow the detection of significantly more candidates, building the first census and opening the way to follow-up spectroscopic observations with the \textit{JWST}. A major role in the search for candidates will be played by \textit{WFIRST}.
Below we review the observational requirements and how they compare with planned instruments (see Fig. \ref{fig:schematic}).

\textbf{Near Infrared:} Employing radiation-hydrodynamic simulations, several studies have shown that the peak emission for a typical heavy seed at $z \gtrsim 10$ falls in the near-infrared, at $\approx 1 \, \mathrm{\mu m}$ \citep{PFVD_2015, Inayoshi_2016, Natarajan_2017, Valiante_2018_b}. This fact is true under very general conditions, i.e. independently of the time elapsed from the formation of the seed, of the environmental metallicity and of the specifics of the accretion. The same studies show that the luminosity of a typical heavy seed varies wildly depending on the aforementioned parameters. It is very challenging to define the ``expected luminosity'' of a high-$z$ seed, as we are dealing with massively active, and thus rapidly varying, sources. Overall, to observe a $10^5 \, \mathrm{\Msun}$ heavy seed at $z \gtrsim 10$ in the middle of its lifespan ($\sim 100 \, \mathrm{Myr}$) and assuming a Compton-thin scenario, we need to reach a flux density of at least $10^{-16} \, \mathrm{erg \, s^{-1} \, cm^{-2}}$. 
This depth is very well achieved by the planned specifications of \textit{JWST}, expected to reach a flux density limit $\approx 3 \times 10^{-17} \, \mathrm{erg \, s^{-1} \, cm^{-2}}$ for a $10 \, \mathrm{ks}$ exposure at $\sim 1 \, \mathrm{\mu m}$. 
Assuming we can observe any DCBHs accreting at $z \gtrsim 10$ producing a flux density $\gtrsim 3 \times 10^{-17} \, \mathrm{erg \, s^{-1} \, cm^{-2}}$, we calculate how many sources of this kind we expect to observe with \textit{JWST}. The predicted number density of DCBHs widely varies in the range $\sim 10^{-10} - 10^{-1} \, \mathrm{Mpc^{-3}}$ at $z \sim 10$ \citep{Habouzit_2016}. The extreme span in the predictions is mainly due to the uncertainties on the critical external Lyman-Werner radiation field strength that can fully suppress the $H_2$ formation. Employing intermediate values for the number density \citep{Agarwal_2012, Habouzit_2016}, we obtain a number of DCBHs detectable with \textit{JWST} at $z \gtrsim 10$ of $1-10 \, \mathrm{deg^{-2}}$. Assuming more optimistic values for the number density of DCBHs, this prediction would increase by several orders of magnitude.

\textbf{X-rays:} High-energy spectral predictions for typical heavy seeds \citep{PFVD_2015, Valiante_2018_b} need to take into account the fundamental variable of the column density of the host. In general, the X-ray emission is predicted to be bell-shaped, with a peak at $\sim 1 \, \mathrm{keV}$ in the observed frame at $z \sim 10$ and rapidly fading at $\lesssim 0.1 \, \mathrm{keV}$. For Compton-thin sources the minimum requirement to observe a $z \gtrsim 10$ DCBH is a flux density $10^{-16} \, \mathrm{erg \, s^{-1} \, cm^{-2}}$, \citep{Valiante_2018_b} which is well contemplated by the flux limit of \textit{Athena} \citep{Aird_2013}.
For Compton-thick sources ($N_H \gtrsim 1.5 \times 10^{24} \, \mathrm{cm^{-2}}$) instead, more sensitive instruments will be needed, able to reach sources at least $1-2$ orders of magnitude fainter \citep{PFVD_2015}. Flux limits of $\sim10^{-19} \, \mathrm{erg \, s^{-1} \, cm^{-2}}$ are expected to be provided by \textit{Lynx} \citep{Lynx_2018}. In fact, future X-ray missions will play a key role in unraveling the dawn of black holes, especially its most obscured components. Employing again intermediate values for the number density of DCBHs, with \textit{Lynx} we expect to observe $5-50 \, \mathrm{deg^{-2}}$ sources of this kind at $z \gtrsim 10$. As already mentioned in the infrared case, larger values for the number density of DCBHs would lead to a significantly higher prediction for the number of detectable sources.
X-ray and infrared observations are fully synergetic in the study of the dawn of black holes. While infrared bands are invaluable to study the spectral energy distribution and emission lines of high-$z$ sources, X-ray observations will provide crucial insights into their emission mechanisms. Consequently, a high-sensitivity observatory as \textit{JWST} needs to be complemented with equally sensitive high-energy observatories (e.g., \textit{Athena} and \textit{Lynx}) to efficiently study the dawn of black holes. Moreover, a high-sensitivity instrument such as \textit{Lynx} could be able to detect heavy seeds at slightly lower mass scales ($\sim 10^5 \, \mathrm{\Msun}$) when compared to \textit{JWST} \citep{PFVD_2015}, and for a longer period of time in the evolution of the seed (see Fig. \ref{fig:spectrum}, right panel).

\textbf{Gravitational Waves:} Gravitational waves observations will be fundamental to gain a clear view of the population of black hole seeds at $z \gtrsim 10$. According to its technical specifications (e.g. \citealt{Klein_2016}), \textit{LISA} will be able to detect the merger of $\sim 10^5 \, \mathrm{\Msun}$ black hole seeds at $z \gtrsim 8$ with a signal-to-noise ratio of $\sim 200$. Latest predictions \citep{Ricarte_2018} suggest that \textit{LISA} should be able to detect the merger of  $\sim 10^5 \, \mathrm{\Msun}$ heavy seeds in the number of $2-20$ in about 4 years of operations (see also e.g. \citealt{Sesana_2004, Sesana_2007, Tanaka_2009}). Although gravitational waves observations are invaluable to discriminate between seeding models, only synergetic observations in the electromagnetic realm (e.g., with \textit{JWST}, \textit{Lynx}, \textit{Athena}) will provide crucial insights on their formation and growth processes.

\vspace{0.2cm}
\textbf{To conclude, detecting heavy black hole seeds at $z \gtrsim 10$ in the next decade will be challenging but, according to current theoretical models, feasible with upcoming and/or proposed facilities. Their detection will be fundamental to understand the early history of the Universe, as well as its evolution until now.}

\pagebreak

\bibliography{ms}

\end{document}